\documentclass[twocolumn,pra,showpacs,superscriptaddress]{revtex4}
\usepackage[T1]{fontenc}
\usepackage{color}
\usepackage{amsmath}
\usepackage{graphicx}
\usepackage{esint}
\usepackage[unicode=true,pdfusetitle,
 bookmarks=true,bookmarksnumbered=false,bookmarksopen=false,
 breaklinks=false,pdfborder={0 0 1},backref=false,colorlinks=true]
 {hyperref}
\hypersetup{
 citecolor=blue}
\usepackage{breakurl}

\makeatletter
\@ifundefined{textcolor}{}
{%
 \definecolor{BLACK}{gray}{0}
 \definecolor{WHITE}{gray}{1}
 \definecolor{RED}{rgb}{1,0,0}
 \definecolor{GREEN}{rgb}{0,1,0}
 \definecolor{BLUE}{rgb}{0,0,1}
 \definecolor{CYAN}{cmyk}{1,0,0,0}
 \definecolor{MAGENTA}{cmyk}{0,1,0,0}
 \definecolor{YELLOW}{cmyk}{0,0,1,0}
}

\makeatother

\begin{document}

\title{Synchro-thermalization of composite quantum system}

\author{Sheng-Wen Li}

\affiliation{Beijing Computational Science Research Center, Beijing 100084, China}

\author{D. Z. Xu}

\affiliation{Beijing Computational Science Research Center, Beijing 100084, China}

\affiliation{State Key Laboratory of Theoretical Physics, Institute of Theoretical
Physics, the Chinese Academy of Science and University of the Chinese
Academy of Sciences, Beijing 100190, China}

\author{X. F. Liu}

\affiliation{Department of Mathematics, Peking University, Beijing 100871, China}

\author{C. P. Sun}

\affiliation{Beijing Computational Science Research Center, Beijing 100084, China}

\affiliation{Synergetic Innovation Center of Quantum Information and Quantum Physics,
University of Science and Technology of China, Hefei, Anhui 230026,
China}
\begin{abstract}
We study the thermalization of a composite quantum system consisting
of several subsystems, where only a small one of the subsystem contacts
with a heat bath in equilibrium, while the rest of the composite system
is contact free. We show that the whole composite system still can
be thermalized after a relaxation time long enough, if the energy
level structure of the composite system is\emph{ connected}, which
means any two energy levels of the composite system can be connected
by direct or indirect quantum transitions. With an example where an
multi-level system interacts with a set of harmonic oscillators via non-demolition
coupling, we find that the speed of relaxation to the global thermal
state is suppressed by the multi-Franck-Condon factor due to the displacements
of the Fock states when the degrees of freedom is large.
\end{abstract}


\maketitle

\section{Introduction}

Isolated quantum systems evolve unitarily according to the Schr\"odinger
equation. When the quantum system is immersed in a canonical heat
bath with a temperature $T$, after a relaxation time long enough,
it would forget all the initial state information and achieve a canonical
state due to the weak interaction with the environment. This process
is called canonical thermalization \cite{landau_statistical_1980}.

\begin{figure}
\includegraphics[width=1\columnwidth]{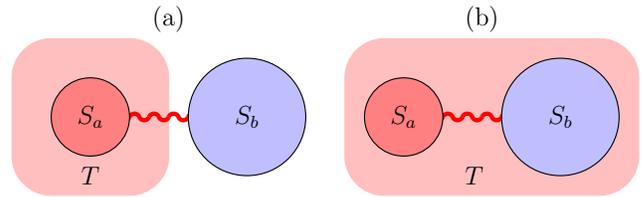}

\caption{(Colored online) Demonstration of a composite system consisting two
interacting parts. (a) Only $S_{a}$ contacts with a heat bath with
temperature $T$, while $S_{b}$ does not. (b) Both parts contact
with the heat bath directly.}
\label{fig-demo}
\end{figure}

Now we consider the thermalization process for a composite system,
which contains several subsystems, where only a small subsystem is
contacted with environment {[}Fig.\,\ref{fig-demo}(a){]}. We show
that if the energy level structure of the composite system is \emph{connected},
i.e., there exists direct or indirect quantum transitions between
any two levels of the system, the whole system would also be thermalized
to its canonical thermal state with the bath temperature $T$. The
contact of a small part would lead to the global thermalization of
the whole system, which is the same as the case that all the subsystems
contact with the same heat bath. We call this process \emph{synchro-thermalization.
}Especially, we consider the case that the interactions between each
subsystems are non-demolition type, where the interaction coupling
does not change the energy of one of the subsystems \cite{dong_quantum_2007,dong_thermodynamic_2009}.
We will explicitly give the condition when the whole composite system
can be synchro-thermalized.

On the first sight, this result seems rather counter-intuitive, especially
when we consider the case that the rest part of the composite system,
which does not contact with the environment, may be quite large and
even tend to be infinite. To solve this puzzle, we consider an example
where an $N$-level system interacts with a set of $M$ harmonic oscillators
via non-demolition coupling. We show that the relaxation rate is suppressed
by a multi-Franck-Condon factor \cite{franck_elementary_1926,condon_theory_1926,lax_franckcondon_1952}:
when the degrees of freedom become large, the speed of global relaxation
to the thermal state would tend to be zero. That means, the larger
the composite system is, the harder it is to thermalize it. And we
call this effect the Franck-Condon blockade \cite{koch_franck-condon_2005}
of thermalization.

We arrange our paper as follows. In Sec.\,II, we give a review on
the general situation of thermalization. In Sec.\,III, we show the
connectivity and thermalization of composite systems. In Sec.\,IV,
we discuss a specific model where the composite system consists an
$N$-level system and an harmonic oscillator. In Sec.\,V, we show
how the thermalization rate is blockaded by the Franck-Condon factor
when the scale of the subsystem free of contact become large. Finally
we draw summary in Sec.\,VI.

\section{Thermalization of a general system}

\emph{}

In order to conveniently use the notations in discussions about the
synchro-thermalization for a composite system, we review some previous
result about the thermalization of a general system first. We write
down the Hamiltonian for the system in the eigen basis $\{|n\rangle\}$
as follows, 
\begin{equation}
\hat{H}_{S}=\sum_{n}\varepsilon_{n}|n\rangle\langle n|,
\end{equation}
where $\varepsilon_{n}$ is the corresponding eigen energy.

The system is contacted with a heat bath via the interaction
\begin{equation}
\hat{H}_{SB}=\sum_{\alpha}\hat{A}_{\alpha}\otimes\hat{R}_{\alpha},
\end{equation}
where $\hat{A}_{\alpha}$ and $\hat{R}_{\alpha}$ are operators of
the system and the heat bath. In the interaction picture of $\hat{H}_{S}+\hat{H}_{B}$,
the system operator $\hat{A}_{\alpha}(t)=\sum_{\omega}\tilde{A}_{\alpha}(\omega)e^{-i\omega t}$
is decomposed according to the oscillation frequencies, and
\begin{equation}
\tilde{A}_{\alpha}(\omega)=\sum_{\varepsilon_{n}-\varepsilon_{m}=\omega}\langle m|\hat{A}_{\alpha}|n\rangle|m\rangle\langle n|
\end{equation}
corresponds to the spectral decomposition of system operator $\hat{A}_{\alpha}(t)$
with respect to the eigen basis $\{|n\rangle\}$.

With the Born-Markovian approximation, the master equation for this
composite system \cite{breuer_theory_2002}, 
\begin{align*}
\dot{\rho}= & \sum_{\omega,\alpha\beta}\gamma_{\alpha\beta}(\omega)\Big(\tilde{A}_{\beta}(\omega)\rho\tilde{A}_{\alpha}^{\dagger}(\omega)-\frac{1}{2}\{\tilde{A}_{\alpha}^{\dagger}(\omega)\tilde{A}_{\beta}(\omega),\,\rho\}_{+}\Big),
\end{align*}
is used to describe the open system dynamics. Here we omitted the
Lamb shift term which has no effect on the thermalization. The generalized
relaxation rates, 
\begin{equation}
\gamma_{\alpha\beta}(\omega)=\int_{-\infty}^{+\infty}d\tau\, e^{i\omega\tau}\langle\hat{R}_{\alpha}^{\dagger}(\tau)\hat{R}_{\beta}(0)\rangle,
\end{equation}
are defined by the bath correlation functions $\langle\hat{R}_{\alpha}^{\dagger}(\tau)\hat{R}_{\beta}(0)\rangle$.

According to the Born approximation, the heat bath keeps unchanged,
i.e., $\rho_{B}(t)\simeq{\cal Z}_{B}^{-1}\exp[-\beta_{T}\hat{H}_{B}]$.
Thus the above finite temperature bath correlation functions $\langle\hat{R}_{\alpha}^{\dagger}(\tau)\hat{R}_{\beta}(0)\rangle$
satisfy the Kubo-Martin-Schwinger condition \cite{kubo_statistical-mechanical_1957,martin_theory_1959}
\begin{equation}
\langle\hat{R}_{\alpha}^{\dagger}(\tau)\hat{R}_{\beta}(0)\rangle=\langle\hat{R}_{\alpha}(0)\hat{R}_{\beta}^{\dagger}(\tau+i\beta_{T})\rangle.
\end{equation}
 It is noticed that the generalized relaxation rates $\gamma_{\alpha\beta}(\omega)$
satisfy the following relation, 
\begin{equation}
\gamma_{\alpha\beta}(-\omega)=e^{-\beta_{T}\omega}\gamma_{\beta\alpha}(\omega),\label{eq:Rate}
\end{equation}
 which is the key point to lead to detailed balance equilibrium \cite{le_bellac_equilibrium_2004}.

The dynamics of the population of the system, denoted as $P_{n}(t):=\rho_{n,n}(t)$,
is decoupled from that of the off-diagonal terms $\rho_{n,m}(t)$,
and can be described by the following Pauli master equation,

\begin{equation}
\dot{P}_{n}=\sum_{m}W(n\leftarrow m)P_{m}-W(m\leftarrow n)P_{n},\label{eq:Pn(t)}
\end{equation}
 where 
\begin{equation}
W(n\leftarrow m)=\sum_{\alpha,\beta}\gamma_{\alpha\beta}(\varepsilon_{m}-\varepsilon_{n})\langle m|\hat{A}_{\alpha}|n\rangle\langle n|\hat{A}_{\beta}|m\rangle,
\end{equation}
 and $W(n\leftarrow m)$ is the time-independent probability transition
rate from $|m\rangle$ to $|n\rangle$. It follows from Eq.\,(\ref{eq:Rate})
that the transition rates satisfies, 
\begin{equation}
W(m\leftarrow n)e^{-\beta_{T}\varepsilon_{n}}-W(n\leftarrow m)e^{-\beta_{T}\varepsilon_{m}}=0.\label{eq:ratio}
\end{equation}

The above derivations have been given in many literatures \cite{breuer_theory_2002}.
Here we focus our attention on the transition structure of the energy
levels induced by the coupling to the heat bath. We would show that
whether the whole system can be thermalized to its unique thermal
state is determined by the \emph{connectivity} of the transition structure
of the energy levels.

If $W(n\leftarrow m)\neq0$, there exists direct probability transition
between the two levels $|n\rangle\leftrightarrow|m\rangle$ . When
$W(m\leftarrow n)=0$, though there is no direct transition between
$|n\rangle$ and $|m\rangle$, the indirect transitions still may
happen which is mediated by some other levels $|k_{i}\rangle$, that
is, the probability transition between $|n\rangle$ and $|m\rangle$
can be completed by a mediating path $ $$|n\rangle\leftrightarrow|k_{1}\rangle\leftrightarrow|k_{2}\rangle\leftrightarrow\dots\leftrightarrow|k_{t}\rangle\leftrightarrow|m\rangle$. 

\begin{figure}
\includegraphics[width=1\columnwidth]{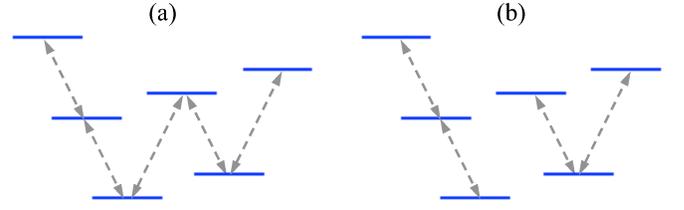}

\caption{(Colored online) Transition structure of the energy levels, (a) connected
(b) disconnected.}
\label{fig-connect}
\end{figure}

Here we use the concept of \emph{connectivity} in topology to describe
such transition structure of the energy levels. We say the two levels
$|n\rangle$ and $|m\rangle$ are \emph{path-connected}, or \emph{connected
by a path}, if and only if there exists a \emph{path}, which is represented
by an ordered series $\{k_{1},\, k_{2},\,\dots,\, k_{t}\}$, such
that $W(n\leftarrow k_{1})\neq0$, $W(m\leftarrow k_{t})\neq0$ and
$W(k_{i}\leftarrow k_{i+1})\neq0$ for any $i$. We say the energy
structure of the whole system is \emph{connected}, if and only if
any two energy levels are path-connected {[}Fig.\,\ref{fig-connect}(a){]},
otherwise, we say the energy structure is \emph{disconnected }{[}Fig.\,\ref{fig-connect}(b){]}\emph{.}

\emph{}

Next, we study the thermalization of the system with the consideration
of topology mentioned above. According to Eq.\,(\ref{eq:ratio}),
the equilibrium steady state requires that \cite{spohn_algebraic_1977,le_bellac_equilibrium_2004,zhang_stochastic_2012}
\begin{align}
 & W(n\leftarrow m)P_{m}-W(m\leftarrow n)P_{n}\nonumber \\
= & W(n\leftarrow m)\Big[P_{m}-e^{-\beta_{T}(\varepsilon_{m}-\varepsilon_{n})}P_{n}\Big]=0.
\end{align}
 If $W(n\leftarrow m)\neq0$, i.e., there exists direct transition
between $|n\rangle$ and $|m\rangle$, so that 
\begin{equation}
P_{n}:P_{m}=e^{-\beta_{T}\varepsilon_{n}}:e^{-\beta_{T}\varepsilon_{m}}.
\end{equation}
If $W(n\leftarrow m)=0$, but $|n\rangle$ and $|m\rangle$ are connected
by a path, denoted as $\{k_{1},\, k_{2},\,\dots,\, k_{t}\}$, we have
$W(n\leftarrow k_{1})\neq0$, $W(m\leftarrow k_{t})\neq0$ and $W(k_{i}\leftarrow k_{i+1})\neq0$.
Thus, with the same reason as above, we have
\begin{align}
P_{n} & :P_{k_{1}}:\dots:P_{k_{t}}:P_{m}\nonumber \\
= & e^{-\beta_{T}\varepsilon_{n}}:e^{-\beta_{T}\varepsilon_{k_{1}}}:\dots:e^{-\beta_{T}\varepsilon_{k_{t}}}:e^{-\beta_{T}\varepsilon_{m}}.\label{eq:proportion}
\end{align}
Therefore, when the energy structure of the system is connected, the
above proportion series includes all the energy levels and that gives
the canonical state
\begin{equation}
\rho=\frac{\sum_{n}e^{-\beta_{T}\varepsilon_{n}}|n\rangle\langle n|}{\sum_{n}e^{-\beta_{T}\varepsilon_{n}}}={\cal Z}^{-1}e^{-\beta_{T}\hat{H}_{S}}.
\end{equation}

When the energy level structure is not connected, but the whole Hilbert
space can be decomposed into two connected subspaces ${\cal V}_{1}$
and ${\cal V}_{2}$ {[}Fig.\,\ref{fig-connect}(b){]}, e.g., spanned
by $\{|n_{1}\rangle\}$ and $\{|n_{2}\rangle\}$ respectively, we
have two independent series, \begin{subequations}
\begin{align}
P_{n_{1}}:P_{m_{1}}:\dots & =e^{-\beta_{T}\varepsilon_{n_{1}}}:e^{-\beta_{T}\varepsilon_{m_{1}}}:\dots,\\
P_{n_{2}}:P_{m_{2}}:\dots & =e^{-\beta_{T}\varepsilon_{n_{2}}}:e^{-\beta_{T}\varepsilon_{m_{2}}}:\dots,
\end{align}
\end{subequations} but we cannot determine further relation between
the two series, which should be determined from the initial state.
Denoting $p_{i}$ as the probability projected into the subspace ${\cal V}_{i}$
from the initial state, it can be verified that 
\begin{equation}
\rho=p_{1}\rho_{1}^{\mathrm{th}}+p_{2}\rho_{2}^{\mathrm{th}}\label{eq:partial-thermal}
\end{equation}
 is the steady state, where 
\begin{equation}
\rho_{i}^{\mathrm{th}}:={\cal Z}_{i}^{-1}\sum_{n_{i}}e^{-\beta_{T}\varepsilon_{n_{i}}}|n_{i}\rangle\langle n_{i}|
\end{equation}
 can be regarded as the partial ``thermal state'' for the connected
subspace ${\cal V}_{i}$.

It follows from Eq.\,(\ref{eq:partial-thermal}) that part of the
initial state information of the system can be preserved if the energy
level structure is not connected. This is different from the connected
case where all the initial state information is erased and the system
is fully thermalized in the steady state.

Now we summarize the above results about the thermalization process
as the following proposition.

\textbf{Proposition:}\emph{ If the energy level structure of the system
is connected, the system can be thermalized to its canonical thermal
state, when it contacts with a equilibrium heat bath. }

\section{Connectivity of composite system and canonical thermalization}

\subsection{Connected case}

Now we study the synchro-thermalization for a composite system, which
consists two subsystems $S_{a}$ and $S_{b}$ coupled with each other.
According to the proposition in the last section, we need to study
whether the energy level structure of the composite system is  connected.
Generally, the Hamiltonian for the whole system is 
\begin{equation}
\hat{H}_{S}=\hat{H}_{a}+\hat{H}_{b}+\hat{V}_{ab},
\end{equation}
 where $\hat{H}_{a}=\sum_{\mathsf{p}}\varepsilon_{\mathsf{p}}^{a}|\mathsf{p}\rangle_{a}\langle\mathsf{p}|$
and $\hat{H}_{b}=\sum_{n}\epsilon_{n}^{b}|n\rangle_{b}\langle n|$.
We denote the eigen state of $\hat{H}_{a}$, $\hat{H}_{b}$ by $\{|\mathsf{p}\rangle_{a}\}$,
$\{|n\rangle_{b}\}$, and $\varepsilon_{\mathsf{p}}^{a}$, $\epsilon_{n}^{b}$
are the corresponding eigen energies. The Hamiltonian of the composite
system $\hat{H}_{S}$ is diagonalized as $\hat{H}_{S}=\sum_{\mathbf{n}}E_{\mathbf{n}}|\mathbf{n}\rangle\langle\mathbf{n}|$,
where $|\mathbf{n}\rangle$ is usually some superposition of the product
states $|\mathsf{p}\rangle_{a}\otimes|n\rangle_{b}$,
\begin{equation}
|\mathbf{n}\rangle=\sum_{\mathsf{p},n}\Psi_{\mathsf{p},n}^{\mathbf{n}}|\mathsf{p}\rangle_{a}|n\rangle_{b}.
\end{equation}
And $E_{\mathbf{n}}$ is the eigen energy of the composite system.

The subsystem $S_{a}$ contacts with a heat bath via the following
interaction, $\hat{H}_{SB}=\sum_{\alpha}\hat{A}_{\alpha}\otimes\hat{R}_{\alpha}$,
where $\hat{A}_{\alpha}$ and $\hat{R}_{\alpha}$ are operators of
$S_{a}$ and the heat bath respectively. We suppose that $S_{b}$
does not contact with any heat bath directly.

$\hat{A}_{\alpha}$ is expanded in the eigen basis of $\hat{H}_{S}$
as $\hat{A}_{\alpha}=\sum_{\mathbf{n},\mathbf{m}}\langle\mathbf{n}|\hat{A}_{\alpha}|\mathbf{m}\rangle\,|\mathbf{n}\rangle\langle\mathbf{m}|$.
Then the transition rate of the composite system becomes
\[
W(\mathbf{n}\leftarrow\mathbf{m})=\sum_{\alpha,\beta}\gamma_{\alpha\beta}(E_{\mathbf{m}}-E_{\mathbf{n}})\langle\mathbf{m}|\hat{A}_{\alpha}|\mathbf{n}\rangle\langle\mathbf{n}|\hat{A}_{\beta}|\mathbf{m}\rangle.
\]

For example, we consider the case that $S_{a}$ is coupled with $S_{b}$
via the interaction of the non-demolition type, i.e., $[\hat{V}_{ab},\hat{H}_{a}]=0$,
which means that the energy of $S_{a}$ does not change due to such
interaction with $S_{b}$ in the absence of the environment. In this
case, the eigen state of the composite system has the form of $|\mathsf{p},n\rangle=|\mathsf{p}\rangle_{a}\otimes|\phi_{n}^{(\mathsf{p})}\rangle_{b}$,
where $|\phi_{n}^{(\mathsf{p})}\rangle_{b}$ is the eigen state of
the effective $\mathsf{p}$-branch Hamiltonian,
\begin{equation}
\hat{H}_{b}^{(\mathsf{p})}:=\hat{H}_{b}+\langle\mathsf{p}|\hat{V}_{ab}|\mathsf{p}\rangle_{a},
\end{equation}
 of the subsystem $S_{b}$. The original energy level $|\mathsf{p}\rangle_{a}$
is splitted into some sub levels $|\mathsf{p},n\rangle$ due to the
interaction. We can write down the transition rates of the composite
system
\begin{align}
W(\mathsf{p}n\leftarrow\mathsf{q}m)= & \sum_{\alpha,\beta}\gamma_{\alpha\beta}(E_{\mathsf{q}m}-E_{\mathsf{p}n})\\
 & \times|\langle\phi_{n}^{(\mathsf{p})}|\phi_{m}^{(\mathsf{q})}\rangle|^{2}\langle\mathsf{p}|\hat{A}_{\alpha}|\mathsf{q}\rangle\langle\mathsf{q}|\hat{A}_{\beta}|\mathsf{p}\rangle.\nonumber 
\end{align}

\emph{}
\begin{figure}
\includegraphics[width=1\columnwidth]{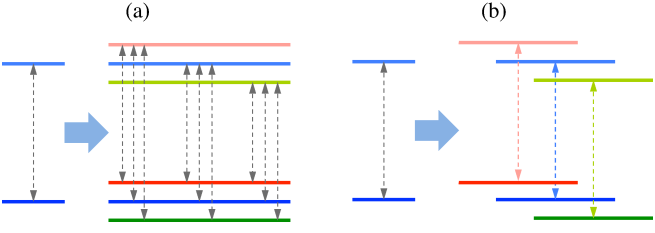}

\caption{(Colored online) Demonstration for the connectivity of the sideband
states of a composite system, (a) connected (b) disconnected.}
\label{fig-sideband}
\end{figure}

In most cases, $W(\mathsf{p}n\leftarrow\mathsf{q}m)$ dose not vanish,
since the matrix elements $\langle\mathbf{m}|\hat{A}_{\alpha}|\mathbf{n}\rangle$
do not vanish simultaneously when the indices $\mathbf{m},\,\mathbf{n},\,\alpha$
take values in their domains. If the original energy levels $|\mathsf{p}\rangle$
of $S_{a}$ are  connected in absence of the interaction with $S_{b}$,
the energy levels of the coupled composite system $|\mathsf{p},n\rangle$
are still  connected {[}Fig.\,\ref{fig-sideband}(a){]}. Therefore,
according to the proposition in the last section, the whole composite
system can be thermalized simultaneously to the canonical state 
\[
\rho_{\mathrm{th}}={\cal Z}^{-1}\exp[-\beta_{T}(\hat{H}_{a}+\hat{H}_{b}+\hat{H}_{ab})].
\]

\subsection{Disconnected case}

Now we present an example where the energy structure of the composite
system is not  connected. We consider the interaction Hamiltonian
$\hat{V}_{ab}$ of non-demolition type for both $S_{a}$ and $S_{b}$,
i.e., $[\hat{V}_{ab},\,\hat{H}_{a/b}]=0$. In this case, $S_{a}$
and $S_{b}$ do not exchange energy with each other, and $\hat{V}_{ab}$
has the form of
\begin{equation}
\hat{V}_{ab}=\sum_{\mathsf{p},n}\, g_{\mathsf{p},n}|\mathsf{p}\rangle_{a}\langle\mathsf{p}|\otimes|n\rangle_{b}\langle n|,
\end{equation}
 and the eigenstates of $\hat{H}_{ab}$ are $|\mathsf{p},n\rangle=|\mathsf{p}\rangle_{a}\otimes|n\rangle_{b}$,
with eigen energy $E_{\mathsf{p}n}=\varepsilon_{\mathsf{p}}^{a}+\epsilon_{n}^{b}+g_{\mathsf{p},n}$.
We obtain the transition rates of the composite system as
\begin{align*}
W(\mathsf{p}n\leftarrow\mathsf{q}m) & =\\
\delta_{mn}\sum_{\alpha,\beta} & \gamma_{\alpha\beta}(E_{\mathsf{q}m}-E_{\mathsf{p}n})\langle\mathsf{p}|\hat{A}_{\alpha}|\mathsf{q}\rangle\langle\mathsf{q}|\hat{A}_{\beta}|\mathsf{p}\rangle.
\end{align*}
Notice that $\gamma_{\alpha\beta}(\omega)$ is usually a smooth function
and varies quite slowly with $\omega$, thus we have $W(\mathsf{p}n\leftarrow\mathsf{q}m)\simeq\delta_{mn}W(\mathsf{p}\leftarrow\mathsf{q})$,
which means that if the states $|\mathsf{p}\rangle$ and $|\mathsf{q}\rangle$
of $S_{a}$ are connected, the sideband states $|\mathsf{p},n\rangle$
and $|\mathsf{q},n\rangle$ are also connected, but $|\mathsf{p},n\rangle$
and $|\mathsf{q},m\rangle$ with $m\neq n$ are not, as demonstrated
in Fig.\,\ref{fig-sideband}(b). As a result, the energy structure
of the composite system is not  connected. According to the discussion
in the last section, the final steady state of the composite system
is $\rho=\sum_{n}p_{n}\rho_{n}^{\mathrm{th}}$, where
\begin{equation}
\rho_{n}^{\mathrm{th}}={\cal Z}_{n}^{-1}\sum_{\mathsf{p}}e^{-\beta_{T}E_{\mathsf{p}n}}|\mathsf{p},n\rangle\langle\mathsf{p},n|,
\end{equation}
 and the probabilities $p_{n}$ are determined by the initial state.
Therefore, the whole composite system cannot be synchro-thermalized
to the canonical state in this situation.

For example, we consider a two-level qubit coupled to a resonator
via the following Hamiltonian,
\begin{align}
\hat{H}_{\mathrm{Q}+\mathrm{L}} & =\hat{H}_{\mathrm{Q}}+\hat{H}_{\mathrm{R}}+\hat{V}_{\mathrm{QR}}\\
 & =-\frac{\varepsilon_{\mathrm{Q}}}{2}\hat{\sigma}^{z}+\omega_{\mathrm{L}}\hat{a}^{\dagger}\hat{a}+g\,\hat{\sigma}^{z}\cdot\hat{a}^{\dagger}\hat{a},\nonumber 
\end{align}
and we have $[\hat{H}_{\mathrm{Q}},\,\hat{V}_{\mathrm{QR}}]=[\hat{H}_{\mathrm{R}},\,\hat{V}_{\mathrm{QR}}]=0$.
This model can be implemented by a Josephson qubit coupled to a superconducting
resonator \cite{lupascu_quantum_2007,schuster_resolving_2007}, or
by an atom inside an optical cavity \cite{brune_quantum_1990}, in
the dispersive regime with large detuning. The resonator is contact
free, while the qubit is contacted with a heat bath via
\[
\hat{H}_{SB}=\hat{\sigma}^{+}\cdot\sum\eta_{k}\hat{b}_{k}+\hat{\sigma}^{-}\cdot\sum\eta_{k}^{*}\hat{b}_{k}^{\dagger}.
\]
The eigen states of $\hat{H}_{\mathrm{Q+L}}$ are $\{|0,n\rangle,\,|1,n\rangle\}$.
We can check that transition happens only between $|0,n\rangle$ and
$|1,n\rangle$, but does not happen between $|\mathsf{p},n\rangle$
and $|\mathsf{q},m\rangle$ for $m\neq n$. The energy structure of
the composite system is disconnected as Fig.\,\ref{fig-sideband}(b).
For this system, the steady state is
\begin{align}
\rho^{\mathrm{th}} & =\sum_{n}p_{n}\rho_{n}^{\mathrm{th}},\\
\rho_{n}^{\mathrm{th}} & =\frac{1}{1+e^{\beta_{T}\varepsilon_{\mathrm{Q}}}}|0,n\rangle\langle0,n|+\frac{e^{\beta_{T}\varepsilon_{\mathrm{Q}}}}{1+e^{\beta_{T}\varepsilon_{\mathrm{Q}}}}|1,n\rangle\langle1,n|,\nonumber 
\end{align}
and $p_{n}$ is determined by the initial condition.

\section{Synchro-thermalization of composite system with non-demolition coupling}

In this section, we consider an example of this synchro-thermalization
of a composite system consisting of an $N$-level system and an harmonic
oscillator interacting with each other via non-demolition coupling.
This composite system is described by the following Hamiltonian,

\begin{align}
\hat{H}_{S} & =\hat{H}_{\mathrm{NLS}}+\hat{H}_{\mathrm{HO}}+\hat{V}_{\mathrm{N-H}}\label{eq:H-N+HO}\\
 & =\sum_{\mathsf{p}=0}^{N-1}\epsilon_{\mathsf{p}}|\mathsf{p}\rangle\langle\mathsf{p}|+\Omega\hat{a}^{\dagger}\hat{a}+\sum_{\mathsf{p}}\xi_{\mathsf{p}}|\mathsf{p}\rangle\langle\mathsf{p}|\,\big(\hat{a}+\hat{a}^{\dagger}).\nonumber 
\end{align}

The $N$-level system is contacted with a heat bath. However, the
harmonic oscillator does not couple to the environment directly. In
the weak coupling limit, the heat bath could be modeled as a collection
of harmonic oscillators linearly coupled to the system \cite{caldeira_quantum_1983}.
Thus the $N$-level system exchanges energy with a boson bath via
the following coupling,
\begin{equation}
\hat{H}_{SB}=\sum_{k}\sum_{\mathsf{p}>\mathsf{q}}^{N-1}g_{k}\big(|\mathsf{p}\rangle\langle\mathsf{q}|\,\hat{b}_{k}+|\mathsf{q}\rangle\langle\mathsf{p}|\,\hat{b}_{k}^{\dagger}),\label{eq:H_SB}
\end{equation}
while the bath Hamiltonian reads $\hat{H}_{B}=\sum_{k}\omega_{k}\hat{b}_{k}^{\dagger}\hat{b}_{k}$.
Here we make a simplification that $g_{k}$ only depend on the bath
mode $k$ but not on the energy level $\mathsf{p},\,\mathsf{q}$,
without loss of generality. 

The Hamiltonian of the composite system Eq.\,(\ref{eq:H-N+HO}) is
diagonalized with the help of $N$ sets of displaced harmonic oscillator
operators corresponding to the coupling with each energy level $|\mathsf{p}\rangle$
\cite{dong_quantum_2007,dong_thermodynamic_2009},
\begin{equation}
\hat{H}_{S}=\sum_{\mathsf{p}=0}^{N-1}|\mathsf{p}\rangle\langle\mathsf{p}|\Big[\Omega\,\hat{D}^{\dagger}(\alpha_{\mathsf{p}})\hat{a}^{\dagger}\hat{a}\hat{D}^{\dagger}(\alpha_{\mathsf{p}})+\varepsilon_{\mathsf{p}}\Big],\label{eq:H-diag}
\end{equation}
 where $\alpha_{\mathsf{p}}=\xi_{\mathsf{p}}/\Omega$ and $\varepsilon_{\mathsf{p}}=\epsilon_{\mathsf{p}}-\xi_{\mathsf{p}}^{2}/\Omega$
are the displacement and energy shift respectively. And we define
the displacement operator as $\hat{D}(\alpha_{\mathsf{p}})=\exp[\alpha_{\mathsf{p}}(\hat{a}^{\dagger}-\hat{a})]$,
which satisfies $\hat{D}^{\dagger}(\alpha_{\mathsf{p}})\hat{a}\hat{D}(\alpha_{\mathsf{p}})=\hat{a}+\alpha_{\mathsf{p}}$.
Then we obtain the eigenstate of $\hat{H}_{S}$ as 
\begin{equation}
|\mathsf{p},n\rangle=|\mathsf{p}\rangle\otimes[\hat{D}^{\dagger}(\alpha_{\mathsf{p}})|n\rangle_{b}],
\end{equation}
 and eigen energy is $E_{\mathsf{p},n}=n\Omega+\varepsilon_{\mathsf{p}}.$
Here $|n\rangle_{b}$ is the Fock state of the original harmonic oscillator.
We notice that each Fock state $|n\rangle_{b}$ is splitted into $N$
levels $|\mathsf{p},n\rangle$ and forms a side-band structure caused
by the coupling with the $N$-level system.

Under the composite eigen basis $\{|\mathsf{p},n\rangle\}$ of $\hat{H}_{S}$,
the interaction Hamiltonian with the bath $\hat{H}_{SB}$ is expressed
as 
\begin{equation}
\hat{H}_{SB}=\sum_{m,n=0}^{\infty}\sum_{\mathsf{p}>\mathsf{q}}^{N-1}\langle n_{\mathsf{p}}|m_{\mathsf{q}}\rangle\,|\mathsf{p},n\rangle\langle\mathsf{q},m|\big(\sum_{k}g_{k}\hat{b}_{k}\big)+\mathbf{h.c.}\label{eq:H-SB-FC}
\end{equation}
 Here, $|n_{\mathsf{p}}\rangle:=\hat{D}^{\dagger}(\alpha_{\mathsf{p}})|n\rangle_{b}$
is the displaced Fock state according to the coupling with $|\mathsf{p}\rangle$.

\begin{figure}
\includegraphics[width=0.98\columnwidth]{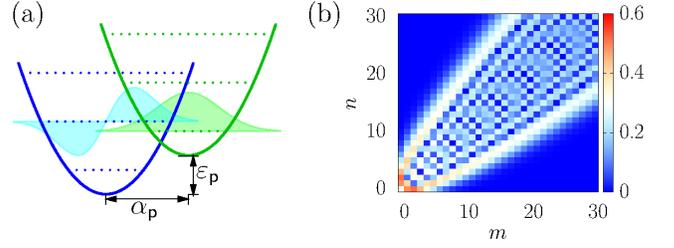}

\caption{(Colored online) (a) Franck-Condon overlap integral for the displaced
Fock states (b) Numerical demonstration for $|\langle n_{\mathsf{p}}|m_{\mathsf{q}}\rangle|=|\langle n|\hat{D}(\alpha_{\mathsf{q}}-\alpha_{\mathsf{p}})|m\rangle|$,
where we set $\alpha_{\mathsf{q}}-\alpha_{\mathsf{p}}=1.5$.}
\label{fig-FC}
\end{figure}

In comparison with the original system-bath coupling Eq.\,(\ref{eq:H_SB}),
the effective coupling strengths of the composite system to the heat
bath are modified by a Franck-Condon factor $\langle n_{\mathsf{p}}|m_{\mathsf{q}}\rangle$,
which is the overlap integral of the wave function of displaced Fock
states \cite{franck_elementary_1926,condon_theory_1926,lax_franckcondon_1952}
(Fig.\,\ref{fig-FC}). As shown as follows, the Franck-Condon factor
suppresses the relaxation rates.

For the boson bath with the coupling spectrum 
\begin{equation}
J(\omega):=2\pi\sum_{k}\left|g_{k}\right|^{2}\delta(\omega-\omega_{k}),\quad\omega\ge0,
\end{equation}
the Born-Markovian approximation gives a master equation for the dynamics
of this composite system \cite{breuer_theory_2002},
\begin{align}
\dot{\rho}=\sum_{\mathsf{p}n,\mathsf{q}m}^{\mathsf{q}\neq\mathsf{p}}\Gamma(\Delta_{\mathsf{p}n,\mathsf{q}m}) & \Big(\hat{L}_{\mathsf{p}n,\mathsf{q}m}\rho\hat{L}_{\mathsf{p}n,\mathsf{q}m}^{\dagger}\nonumber \\
 & -\frac{1}{2}\{\hat{L}_{\mathsf{p}n,\mathsf{q}m}^{\dagger}\hat{L}_{\mathsf{p}n,\mathsf{q}m},\rho\}\Big),\label{eq:ME}
\end{align}
 where $\hat{L}_{\mathsf{p}n,\mathsf{q}m}:=|\mathsf{p}n\rangle\langle\mathsf{q}m|$
is the Lindblad operator, and $\Delta_{\mathsf{p}n,\mathsf{q}m}:=E_{\mathsf{p}n}-E_{\mathsf{q}m}$.
$\Gamma(\Delta_{\mathsf{p}n,\mathsf{q}m})$ is the dissipation rate
between the two levels $|\mathsf{p},n\rangle$ and $|\mathsf{q},m\rangle$,
and
\[
\Gamma_{\mathsf{p}n,\mathsf{q}m}(\omega)=\begin{cases}
\left|\langle n_{\mathsf{p}}|m_{\mathsf{q}}\rangle\right|^{2}J(\omega)N(\omega), & \omega\ge0\\
\left|\langle n_{\mathsf{p}}|m_{\mathsf{q}}\rangle\right|^{2}J(|\omega|)[N(|\omega|)+1], & \omega<0
\end{cases}
\]
The Franck-Condon factor also appears in this dissipation rate. The
norm $|\langle n_{\mathsf{p}}|m_{\mathsf{q}}\rangle|\le1$ gives that
the relaxation rates are suppressed by the Franck-Condon factor (Fig.\,\ref{fig-FC}).

The rate equation about the dynamics of the energy population $P_{\mathsf{p}n}:=\langle\mathsf{p}n|\rho|\mathsf{p}n\rangle$,
\begin{equation}
\dot{P}_{\mathsf{p}n}=\sum_{\mathsf{q},m}^{\mathsf{q}\neq\mathsf{p}}W(\mathsf{p}n\leftarrow\mathsf{q}m)P_{\mathsf{q}m}-W(\mathsf{q}m\leftarrow\mathsf{p}n)P_{\mathsf{p}n},\label{eq:rate-Eq}
\end{equation}
 is obtained from the above master equation, which is decoupled from
that of the off-diagonal terms. Here $W(\mathsf{p}n\leftarrow\mathsf{q}m)=\Gamma(\Delta_{\mathsf{p}n,\mathsf{q}m})$
is the population transition rate. The steady state condition requires
$\dot{P}_{\mathsf{p}n}=0$, which gives, 
\begin{align*}
\left|\langle n_{\mathsf{p}}|m_{\mathsf{q}}\rangle\right|^{2}J(\Delta_{\mathsf{p}n,\mathsf{q}m})\times\\
\Big[N(\Delta_{\mathsf{p}n,\mathsf{q}m})P_{\mathsf{q}m}- & \big(N(\Delta_{\mathsf{p}n,\mathsf{q}m})+1\big)P_{\mathsf{p}n}\Big]=0,
\end{align*}
 for all $E_{\mathsf{p}n}>E_{\mathsf{q}m}$ with $\mathsf{p}\neq\mathsf{q}$.

For any two different levels $|\mathsf{p},n\rangle$ and $|\mathsf{q},m\rangle$,
where $\mathsf{p}\neq\mathsf{q}$, we have $W(\mathsf{p}n\leftarrow\mathsf{q}m)\neq0$.
Thus, under the mediation of the environment, the eigenstates of the
composite system $|\mathsf{p}n\rangle$ are connected. And the steady
population of each two eigenstates satisfies the Boltzmann distribution
\[
P_{\mathsf{p}n}:P_{\mathsf{q}m}=e^{-\beta_{T}E_{\mathsf{p}n}}:e^{-\beta_{T}E_{\mathsf{q}m}}.
\]
 Therefore, the whole composite state can be stabilized to its canonical
thermal state $\rho_{\mathrm{th}}={\cal Z}^{-1}\exp[-\beta(\hat{H}_{\mathrm{NLS}}+\hat{H}_{\mathrm{HO}}+\hat{V}_{\mathrm{N-H}})]$
when $t\rightarrow\infty$. The composite system is synchro-thermalized
as we discussed in Sec.\,III-A.

\emph{}

\section{Franck-Condon blockade}

We have shown that for a composite system, a partial contact with
a heat bath would cause global thermalization. It seems that even
if the subsystem contacted with the heat bath is quite small while
the rest part tends to be infinitely large, the whole system still
can be thermalized globally. In this sense, the above result is rather
counter-intuitive.

In order to solve this puzzle, we consider the subsystem $S_{b}$
consists of $M$ harmonic oscillators. $M$ characterizes the scale
of the subsystem free of the coupling to the bath. We will see that
indeed the global thermalization rate of the whole system decreases
rapidly when the scale of the whole composite system becomes large.

We generalize the above example by using $M$ harmonic oscillators
to replace the single one coupled to the $N$-level system. Then the
system Hamiltonian reads
\begin{align}
\hat{H}_{S}=\sum_{\mathsf{p}=0}^{N-1}\epsilon_{\mathsf{p}}|\mathsf{p}\rangle\langle\mathsf{p}| & +\sum_{i=1}^{M}\Omega_{i}\hat{a}_{i}^{\dagger}\hat{a}_{i}\nonumber \\
 & +\sum_{i,\mathsf{p}}\xi_{i,\mathsf{p}}|\mathsf{p}\rangle\langle\mathsf{p}|\,\big(\hat{a}_{i}+\hat{a}_{i}^{\dagger}),\label{eq:H-N+HOs}
\end{align}
which is diagonalized as 
\begin{equation}
\hat{H}_{S}=\sum_{i,\mathsf{p}}\Big[\Omega_{i}\,\hat{D}^{\dagger}(\alpha_{i,\mathsf{p}})\hat{a}_{i}^{\dagger}\hat{a}_{i}\hat{D}^{\dagger}(\alpha_{i,\mathsf{p}})+\varepsilon_{i,\mathsf{p}}\Big]\,|\mathsf{p}\rangle\langle\mathsf{p}|,\label{eq:Hs-diag}
\end{equation}
where $\alpha_{i,\mathsf{p}}$ and $\varepsilon_{i,\mathsf{p}}$ are
the displacement and energy shift according to the $i$-th harmonic
oscillator and level-$\mathsf{p}$, and $\alpha_{i,\mathsf{p}}:=\xi_{i,\mathsf{p}}/\Omega_{i},\,\varepsilon_{i,\mathsf{p}}:=\epsilon_{i,\mathsf{p}}-\xi_{i,\mathsf{p}}^{2}/\Omega_{i}$.
We obtain the eigenstate and eigen energy of $\hat{H}_{S}$, i.e.,
\begin{align}
|\mathsf{p},\vec{\mathbf{n}}\rangle & =|\mathsf{p}\rangle\otimes\big[\bigotimes_{i=1}^{M}\hat{D}^{\dagger}(\alpha_{i,\mathsf{p}})|n\rangle_{i}\big],\\
E_{\vec{\mathbf{n}},\mathsf{p}} & =\sum_{i}n_{i}\Omega_{i}+\varepsilon_{i,\mathsf{p}}.\nonumber 
\end{align}
where we use $\vec{\mathbf{n}}:=(n_{1},\, n_{2},\,\dots,\, n_{M})$
to denote the quantum numbers of $M$ harmonic oscillators.

\emph{}
\begin{figure}
\includegraphics[width=0.85\columnwidth]{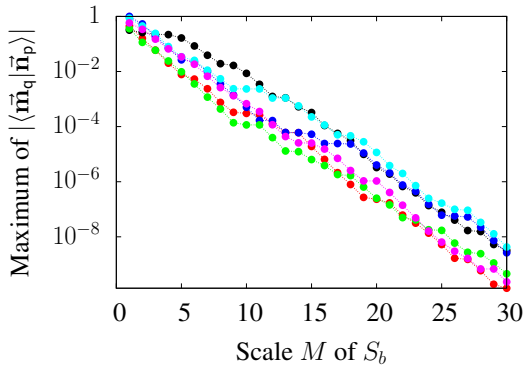}

\caption{(Colored online) Demonstration for multi-Franck-Condon factor $\left|\langle\vec{\mathbf{n}}_{\mathsf{p}}|\vec{\mathbf{m}}_{\mathsf{q}}\rangle\right|$
by product from the maximal matrix elements of $|\langle n_{i}|\hat{D}(\Delta\alpha_{i,\mathsf{qp}}|m_{i}\rangle|$.
The displacements are simulated by $M$ random number $\Delta\alpha_{i,\mathsf{qp}}\in(-4,4)$.
We show 6 groups of samples, which decay exponentially with the system
scale $M$.}
\label{fig-max}
\end{figure}

Correspondingly, the master equation are obtained similarly to Eqs.\,(\ref{eq:ME},\,\ref{eq:rate-Eq}),
and the thermalization rates for this composite system are
\[
\Gamma_{\mathsf{p}\vec{\mathbf{n}},\mathsf{q}\vec{\mathbf{m}}}(\omega)=\begin{cases}
\left|\langle\vec{\mathbf{n}}_{\mathsf{p}}|\vec{\mathbf{m}}_{\mathsf{q}}\rangle\right|^{2}J(\omega)N(\omega), & \omega\ge0,\\
\left|\langle\vec{\mathbf{n}}_{\mathsf{p}}|\vec{\mathbf{m}}_{\mathsf{q}}\rangle\right|^{2}J(|\omega|)[N(|\omega|)+1], & \omega<0.
\end{cases}
\]
Comparing with the above case where $S_{b}$ only contains one harmonic
oscillator, the only difference is that the thermalization rates are
modified with a multi-Franck-Condon factor 
\begin{equation}
\left|\langle\vec{\mathbf{n}}_{\mathsf{p}}|\vec{\mathbf{m}}_{\mathsf{q}}\rangle\right|^{2}=\prod_{i=1}^{M}\left|\langle n_{i,\mathsf{p}}|m_{i,\mathsf{q}}\rangle\right|^{2}.
\end{equation}

Since each term in the product $\left|\langle n_{i,\mathsf{p}}|m_{i,\mathsf{q}}\rangle\right|^{2}<1$,
when $M$ becomes large, the whole factor $\Gamma_{\mathsf{p}\vec{\mathbf{n}},\mathsf{q}\vec{\mathbf{m}}}(\omega)$
tends to vanish, and the thermalization rate approaches infinitesimal.
That means, the speed of relaxation to the global thermal state is
greatly suppressed by the multi-Franck-Condon factor due to the displacements
of the Fock states when the degrees of freedom is large. We call this
effect Franck-Condon blockade \cite{koch_franck-condon_2005}. 

Here we estimate the maximum of the multi-Franck-Condon factor as,
\begin{align*}
\left|\langle\vec{\mathbf{n}}_{\mathsf{p}}|\vec{\mathbf{m}}_{\mathsf{q}}\rangle\right|^{2} & \le\prod_{i=1}^{M}\max_{\{m_{i},n_{i}\}}\Big\{\big|[\hat{D}(\Delta\alpha_{i,\mathsf{qp}})]_{m_{i},n_{i}}\big|^{2}\Big\}\\
 & \sim\exp[-M\langle\Delta\alpha_{i,\mathsf{qp}}^{2}\rangle]
\end{align*}
 where $\Delta\alpha_{i,\mathsf{qp}}:=\alpha_{i,\mathsf{q}}-\alpha_{i,\mathsf{p}}$.
The above inequality achieves $1$ if and only if $\Delta\alpha_{i,\mathsf{qp}}=0$
for any $\mathsf{p},\,\mathsf{q},\, i$, namely, $\xi_{i,\mathsf{p}}\equiv\xi_{i}$
does not depend on $\mathsf{p}$ {[}Eq.\,(\ref{eq:H-N+HOs}){]}.
In this case the $N$-level system is indeed decoupled from the $M$
oscillators, and the interaction term only contributes displacements
to the oscillators. The thermalization speed returns to the case for
the thermalization of the $N$-level system alone, regardless of the
$M$ oscillators. In usual cases, the multi-Franck-Condon factor gives
rise to a exponential suppressing of the thermalization speed, which
decays quite fast with the scale $M$ of the subsystem $S_{b}$ (Fig.\,\ref{fig-max}).

\section{Summary}

In this paper, we have studied the synchro-thermalization for a composite
system consisting of several subsystems, where only one of the subsystems
is contacted with a canonical heat bath. It has been shown that the
partial contact of one subsystem to the heat bath may stabilize the
whole composite system to its canonical thermal state. We clarified
that the conditions for the synchro-thermalization depends on the
topology of the eigen-energy-level structure, i.e., whether it is
 connected or not. We illustrated our main results with an $N$-level
system coupled to several harmonic oscillators via the non-demolition
way. 

Our arguments lead to a puzzle that the whole composite system can
be thermalized even if the scale of the subsystems free of contact
tends to be infinitely large. This puzzle can be explicitly solved
by our illustration with the Franck-Condon blockade mechanism. When
the degrees of freedom of the composite system becomes large, the
thermalization rate would approach zero. That is to say, it becomes
more and more difficult to stabilize the composite system to its canonical
thermal state. Thus this effect of synchro-thermalization is not easy
to be observed for large scale systems.

This work is supported by National Natural Science Foundation of China
under Grants Nos.\,11121403, 10935010 and 11074261, National 973-program
Grants No.\,2012CB922104, and Postdoctoral Science Foundation of
China No.\,2013M530516.

\end{document}